\documentclass[12pt]{article}

\usepackage{amsmath,amssymb,amscd,amsbsy,amsgen,amsopn,amstext,
amsxtra}

\usepackage[dvips]{graphicx}

\begin{document}

\vskip 0.5 truecm

\begin{center}
{\Large{\bf Gauge symmetries in geometric phases\footnote{Talk given at Summer Institute 2005, Fuji-Yoshida, August 11-18, 2005 (to be published in Proceedings).}}}
\end{center}
\vskip .5 truecm
\centerline{\bf  Kazuo Fujikawa }
\vskip .4 truecm
\centerline {\it Institute of Quantum Science, College of 
Science and Technology}
\centerline {\it Nihon University, Chiyoda-ku, Tokyo 101-8308, 
Japan}
\vskip 0.5 truecm

\begin{abstract}
The analysis of geometric phases is briefly reviewed by emphasizing various gauge symmetries involved. The analysis of geometric phases
 associated with level crossing is reduced to the familiar 
diagonalization of the Hamiltonian in the second quantized 
formulation. A hidden local gauge symmetry becomes explicit in 
this formulation and specifies physical observables; the choice 
of a basis set which specifies the coordinates in the functional
 space is arbitrary in the second quantization, and a sub-class 
of coordinate transformations, which keeps the form of the 
action invariant, is recognized as the gauge symmetry. 
It is shown that the hidden local symmetry provides a basic 
concept which replaces the notions of parallel transport and 
holonomy. We also point out that our hidden local gauge symmetry
 is quite different from a gauge symmetry used by Aharonov and 
Anandan in their definition of non-adiabatic phases. 
\end{abstract}

%\makeatletter
%\@addtoreset{equation}{section}
%\def\theequation{\thesection.\arabic{equation}}
%\makeatother
%\large

\section{Introduction}
The conventional machinery to analyze geometric phases is a 
perecise adiabatic approximation and the notions of parallel 
transport and holonomy\cite{simon}. The common less stringent 
but more
flexible approaches are based on the practical Born-Oppenheimer 
approximation where the period $T$ of the slower system in units
 of the time scale of the faster system is 
finite~\cite{berry, stone, berry2}. The idea of the 
non-adiabatic phase, which does not rely on the adiabatic 
approximation, has also been proposed~\cite{aharonov, aitchison,
 mukunda}; the gauge symmetry used in this proposal is 
explained  later.

Here I review  a second quantized formulation of geometric 
phases\cite{fujikawa,fujikawa2,fujikawa3} with an emphasis on 
gauge symmetries involved: The possible advantages of this 
formulation are\\
1. All the phases become purely dynamical, i.e., part of the Hamiltonian, and thus the analysis of geometric phases is reduced 
to a simple diagonalization of the Hamiltonian. This formulation
 also works for both of the path integral and operator 
formulations.\\
2. Origins of various gauge symmetries used in the past analyses
become transparent. In particular, the presence of a hidden 
local gauge symmetry in the analysis of adiabatic phases
is clearly recognized.\\
3. One obtains a better view of topological properties, such as 
the absence of the monopole-like singularity.\\
4. Differences between the quantum anomaly and the geometric 
phase become clear in this formulation.
\\
 
From the point of view of particle physicists, the essence of the
geometric phase may be summarized as follows: The essence of the 
geometric phase is to use an {\em approximation} (adiabatic 
approximation) to obtain a clear universal view of what is going 
on in the general level crossing problem, which is not clearly
seen in the exact treatment. Mathematically, a time dependent
unitary transformation which is singular at the level crossing
point palys a central role.

\section{Second quantized formulation and geometric phases}

We start with the generic (hermitian) Hamiltonian 
\begin{eqnarray}
\hat{H}=\hat{H}(\hat{\vec{p}},\hat{\vec{x}},X(t))
\end{eqnarray}
for a single particle theory in a slowly varying background 
variable \\
$X(t)=(X_{1}(t),X_{2}(t),...)$.
The path integral for this theory for the time interval
$0\leq t\leq T$ in the second quantized 
formulation is given by 
\begin{eqnarray}
&&\int{\cal D}\psi^{\star}{\cal D}\psi
\exp\{\frac{i}{\hbar}\int_{0}^{T}dtd^{3}x[{\cal L}
] \},
\nonumber\\
{\cal L}&=&\psi^{\star}(t,\vec{x})i\hbar\frac{\partial}{\partial t}
\psi(t,\vec{x})-\psi^{\star}(t,\vec{x})
\hat{H}(\frac{\hbar}{i}\frac{\partial}{\partial\vec{x}},
\vec{x},X(t))\psi(t,\vec{x})
\end{eqnarray}
We then define a complete set of eigenfunctions
\begin{eqnarray}
&&\hat{H}(\frac{\hbar}{i}\frac{\partial}{\partial\vec{x}},
\vec{x},X(0))u_{n}(\vec{x},X(0))=\lambda_{n}u_{n}(\vec{x},X(0)), \nonumber\\
&&\int d^{3}xu_{n}^{\star}(\vec{x},X(0))u_{m}(\vec{x},X(0))=
\delta_{nm},
\end{eqnarray}
and expand 
\begin{eqnarray}
\psi(t,\vec{x})=\sum_{n}a_{n}(t)u_{n}(\vec{x},X(0)).
\end{eqnarray}
We then have
\begin{eqnarray}
{\cal D}\psi^{\star}{\cal D}\psi=\prod_{n}{\cal D}a_{n}^{\star}
{\cal D}a_{n}
\end{eqnarray}
and the path integral is written as 
\begin{eqnarray}
Z&=&\int \prod_{n}{\cal D}a_{n}^{\star}
{\cal D}a_{n}
\exp\{\frac{i}{\hbar}\int_{0}^{T}dt{\cal L} \},\nonumber\\
{\cal L}&=&
\sum_{n}a_{n}^{\star}(t)i\hbar\frac{\partial}{\partial t}
a_{n}(t)-\sum_{n,m}a_{n}^{\star}(t)E_{nm}(X(t))a_{m}(t)
\end{eqnarray}
where 
\begin{eqnarray}
&&E_{nm}(X(t))=\int d^{3}x u_{n}^{\star}(\vec{x},X(0))
\hat{H}(\frac{\hbar}{i}\frac{\partial}{\partial\vec{x}},
\vec{x},X(t))u_{m}(\vec{x},X(0)).
\end{eqnarray}

We next perform a unitary transformation
\begin{eqnarray}
a_{n}(t)=\sum_{m}U(X(t))_{nm}b_{m}(t)
\end{eqnarray}
where 
\begin{eqnarray}
&&U(X(t))_{nm}=\int d^{3}x u^{\star}_{n}(\vec{x},X(0))
v_{m}(\vec{x},X(t))
\end{eqnarray}
with the instantaneous eigenfunctions of the Hamiltonian
\begin{eqnarray}
&&\hat{H}(\frac{\hbar}{i}\frac{\partial}{\partial\vec{x}},
\vec{x},X(t))v_{n}(\vec{x},X(t))={\cal E}_{n}(X(t))v_{n}(\vec{x},X(t)), \nonumber\\
&&\int d^{3}x v^{\star}_{n}(\vec{x},X(t))v_{m}(\vec{x},X(t))
=\delta_{n,m}.
\end{eqnarray}
We take the time $T$ 
as a period of the slowly varying variable $X(t)$ in the analysis
of geometric phases.

We can thus re-write the path integral as 
\begin{eqnarray}
&&Z=\int \prod_{n}{\cal D}b_{n}^{\star}{\cal D}b_{n}
\exp\{\frac{i}{\hbar}\int_{0}^{T}dt{\cal L} \},\nonumber\\
&&{\cal L}=
\sum_{n}b_{n}^{\star}(t)i\hbar\frac{\partial}{\partial t}
b_{n}(t)-\sum_{n}b_{n}^{\star}(t){\cal E}_{n}(X(t))b_{n}(t)
\nonumber\\
&&+\sum_{n,m}b_{n}^{\star}(t)
\langle n|i\hbar\frac{\partial}{\partial t}|m\rangle
b_{m}(t)
\end{eqnarray}
where the last term in the action stands for the term
commonly referred to as Berry's phase and its 
off-diagonal generalization. 
This term is defined by
\begin{eqnarray} 
(U(X(t))^{\dagger}i\hbar\frac{\partial}{\partial t}
U(X(t)))_{nm}
&=&\int d^{3}x v^{\star}_{n}(\vec{x},X(t))
i\hbar\frac{\partial}{\partial t}v_{m}(\vec{x},X(t))\nonumber\\
&\equiv& \langle n|i\hbar\frac{\partial}{\partial t}|m\rangle.
\end{eqnarray}
The use of the instantaneous eigenfunctions here is
 a common feature shared with the adiabatic approximation. In 
our picture, all the information about geometric phases  is 
included in the effective Hamiltonian and thus purely 
{\em dynamical}.

When one defines the Schr\"{o}dinger picture by
\begin{eqnarray}
\hat{{\cal H}}_{eff}(t)&\equiv&
\sum_{n}\hat{b}_{n}^{\dagger}(0){\cal E}_{n}(X(t))\hat{b}_{n}(0)
-\sum_{n,m}\hat{b}_{n}^{\dagger}(0)
\langle n|i\hbar\frac{\partial}{\partial t}|m\rangle
\hat{b}_{m}(0)
\end{eqnarray}
in the operator formulation, the second quantized formula for
 the evolution operator gives
\begin{eqnarray}
&&\langle m|T^{\star}\exp\{-\frac{i}{\hbar}\int_{0}^{T}
\hat{{\cal H}}_{eff}(t)
dt\}|n\rangle\nonumber\\
&&=
\langle m(T)|T^{\star}\exp\{-\frac{i}{\hbar}\int_{0}^{T}
\hat{H}(\hat{\vec{p}}, \hat{\vec{x}},  
X(t))dt \}|n(0)\rangle
\end{eqnarray}
where $T^{\star}$ stands for the time ordering operation, and 
the state vectors in the second quantization  on the left-hand 
side are defined by 
\begin{eqnarray}
|n\rangle=\hat{b}_{n}^{\dagger}(0)|0\rangle,
\end{eqnarray} 
and the state vectors on the right-hand side  stand for the 
first quantized states defined by
\begin{eqnarray}
\langle \vec{x}|n(t)\rangle=v_{n}(\vec{x},X(t))
\end{eqnarray}
appearing in (10).
Both-hand sides of the above equality (14) are exact, but the 
difference is that the geometric terms, both of diagonal and 
off-diagonal, are explicit in the second quantized formulation 
on the left-hand side.

The Schr\"{o}dinger amplitude is given by
\begin{eqnarray}
\psi_{n}(\vec{x},T; X(T))=
\langle 0|\hat{\psi}(T,\vec{x})\hat{b}^{\dagger}_{n}(0)|0\rangle.
\end{eqnarray}
In the adiabatic approximation, we assume the dominance of 
diagonal elements
\begin{eqnarray}
\psi_{n}(\vec{x},T; X(T))
&=&\sum_{m}v_{m}(\vec{x};X(T))
\langle m|T^{\star}\exp\{-\frac{i}{\hbar}\int_{0}^{T}
\hat{{\cal H}}_{eff}(t)dt\}|n\rangle\nonumber\\
&\simeq& v_{n}(\vec{x};X(T))
\exp\{-\frac{i}{\hbar}\int_{0}^{T}[{\cal E}_{n}(X(t))
-\langle n|i\hbar\frac{\partial}{\partial t}|n\rangle]dt\}.
\end{eqnarray}
 
\section{ Hidden local gauge symmetry}

The path integral formula (11) is 
based on the expansion
\begin{eqnarray}
\psi(t,\vec{x})=\sum_{n}b_{n}(t)v_{n}(\vec{x},X(t)),
\end{eqnarray}
and the starting path integral (2) depends only on the field
variable $\psi(t,\vec{x})$, not on  $\{ b_{n}(t)\}$
and $\{v_{n}(\vec{x},X(t))\}$ separately. This fact shows that 
our formulation contains a hidden local gauge symmetry 
\begin{eqnarray}
&&v_{n}(\vec{x},X(t))\rightarrow v^{\prime}_{n}(t; \vec{x},X(t))=
e^{i\alpha_{n}(t)}v_{n}(\vec{x},X(t)),\nonumber\\
&&b_{n}(t) \rightarrow b^{\prime}_{n}(t)=
e^{-i\alpha_{n}(t)}b_{n}(t), \ \ n=1,2,3,.,
\end{eqnarray}
where the gauge parameter $\alpha_{n}(t)$ is a general 
function of $t$. We call this symmetry 
"hidden local gauge symmetry" because it
appears due to the separation of the fundamental dynamical
variable $\psi(t,\vec{x})$ into two sets $\{ b_{n}(t)\}$
and $\{v_{n}(\vec{x},X(t))\}$.  
One can confirm that the action and the path integral measure in
(11) are both invariant under this gauge transformation. 

The above hidden local symmetry is exact as long as the 
basis set is not singular. In the present problem, the basis
set  becomes singular on top of level 
crossing, and thus the above symmetry is 
particularly useful in the general adiabatic approximation 
defined by the condition 
that the basis set is well-defined. Of course, one may  
consider a new hidden local gauge symmetry when one defines a new
regular coordinate in the neighborhood of the singularity, and 
the freedom in the phase choice of the new basis set persists.
Physically, this hidden gauge symmetry arises from the 
fact  that the 
choice of the basis set which specifies the coordinates in the 
functional space is arbitrary in field theory, as long as the 
coordinates are not singular.

In practical applications for  
generic eigenvalues $\{{\cal E}_{n}(X(t))\}$, the sub-group
\begin{eqnarray}
U=U(1)\times U(1)\times .....
\end{eqnarray}
as in the above (20) is useful, because it keeps the form of the action 
invariant and thus becomes a symmetry of quantized theory in the
 conventional sense. In particular, it is exactly 
preserved in the adiabatic approximation in which the mixing of
different energy eigenstates is assumed to be negligible and thus
the coordinates specified are always well-defined. 
 
 For a special case where the first eigenvalue
${\cal E}_{1}(X(t))$ has $n_{1}$-fold degeneracy, the second
eigenvalue ${\cal E}_{2}(X(t))$ has $n_{2}$-fold degeneracy, and 
so on, the sub-group
\begin{eqnarray}
U=U(n_{1})\times U(n_{2})\times ....,
\end{eqnarray}
which keeps the form of the action invariant, will be  useful.

The above hidden local gauge 
symmetry is an exact symmetry of quantum theory, and thus 
physical observables in the adiabatic approximation should 
respect this symmetry. Also, by using this local gauge freedom, 
one can choose the phase convention of the basis set
$\{v_{n}(t,\vec{x},X(t))\}$ such that the analysis of geometric 
phases becomes most transparent.

Under the hidden local symmetry, the probability amplitude in
(17) transforms 
\begin{eqnarray}
&&\psi^{\prime}_{n}(\vec{x},t; X(t))=e^{i\alpha_{n}(0)}
\psi_{n}(\vec{x},t; X(t))
\end{eqnarray}
{\em independently} of the value of $t$. 
Thus the product
\begin{eqnarray}
\psi_{n}(\vec{x},0; X(0))^{\star}\psi_{n}(\vec{x},T; X(T))
\end{eqnarray}
defines a manifestly gauge invariant quantity, namely, it is 
independent of the choice of the phase convention of the 
complete basis set $\{v_{n}(t,\vec{x},X(t))\}$.

For the adiabatic formula (18), the gauge invariant quantity is given
 by
\begin{eqnarray}
\psi_{n}(\vec{x},0; X(0))^{\star}\psi_{n}(\vec{x},T; X(T))
&=&v_{n}(0,\vec{x}; X(0))^{\star}v_{n}(T,\vec{x};X(T))\\
&&\times\exp\{-\frac{i}{\hbar}\int_{0}^{T}[{\cal E}_{n}(X(t))
-\langle n|i\hbar\frac{\partial}{\partial t}|n\rangle]dt\}
\nonumber
\end{eqnarray}
and the combination 
\begin{eqnarray}
v_{n}(0,\vec{x}; X(0))^{\star}v_{n}(T,\vec{x};X(T))\exp\{-\frac{i}{\hbar}\int_{0}^{T}[
-\langle n|i\hbar\frac{\partial}{\partial t}|n\rangle]dt\}
\end{eqnarray}
is invariant under the hidden local gauge symmetry. By choosing 
the gauge such that 
\begin{eqnarray}
v_{n}(T,\vec{x};X(T))=v_{n}(0,\vec{x}; X(0))
\end{eqnarray}
the prefacotor
$v_{n}(0,\vec{x}; X(0))^{\star}v_{n}(T,\vec{x};X(T))$ becomes 
real and positive. Note that we are assuming the cyclic motion of
the external parameter, $X(T)=X(0)$.
Then the factor
\begin{eqnarray}
\exp\{-\frac{i}{\hbar}\int_{0}^{T}[{\cal E}_{n}(X(t))
-\langle n|i\hbar\frac{\partial}{\partial t}|n\rangle]dt\}
\end{eqnarray}
extracts all the information about the phase in (25) and 
defines a physical quantity. After this gauge fixing, the 
above quantity is still invariant under residual gauge 
transformations satisfying the periodic boundary condition
\begin{eqnarray}
\alpha_{n}(0)=\alpha_{n}(T),
\end{eqnarray}
in particular, for a class of gauge transformations defined 
by $\alpha_{n}(X(t))$. Note that our gauge transformation, which
 is defined by an arbitrary function 
$\alpha_{n}(t)$,  is more general. 
It has been shown that this hidden gauge symmetry replaces the 
notions of parallel transport and holonomy in the analysis of 
geometric phases. We note that the notion such as holonomy is 
valid only in the limit of the very precise adiabatic approximation~\cite{simon}.

\section{ Explicit example; two-level truncation}
 
We analyze the crossing of two levels in the above general model.
In the sufficiently close to the level crossing point, we assume
that we can truncate the model to a simplified two-level model
which contains the two levels at issue.
The effective Hamiltonian for the Lagarangian in (6) is then 
reduced to the 
$2\times 2$ matrix $ h(X(t))=\left(E_{nm}(X(t))\right)$.
If one assumes that the level crossing takes place at the 
origin of the parameter space $X(t)=0$, one  analyzes
the matrix
\begin{eqnarray}
h(X(t)) = \left(E_{nm}(0)\right)+\left(\frac{\partial}{\partial X_{k}}E_{nm}(X)|_{X=0}\right)
X_{k}(t)
\end{eqnarray}
 for sufficiently small $(X_{1}(1),X_{2}(1), ... )$.  After a 
suitable definition of the parameters $y(t)$ by taking linear 
combinations of  $X_{k}(t)$, we write the matrix 
\begin{eqnarray}
h(X(t))
&=&\left(\begin{array}{cc}
            E(0)+y_{0}(t)&0\\
            0&E(0)+y_{0}(t)
            \end{array}\right) +g \sigma^{l}y_{l}(t)
\end{eqnarray}
where $\sigma^{l}$ stands for the Pauli matrices, and $g$ is a 
suitable (positive) coupling constant.
 
The above matrix is diagonalized in the standard way as 
\begin{eqnarray} 
h(X(t))v_{\pm}(y)=(E(0)+y_{0}(t) \pm g r)v_{\pm}(y)
\end{eqnarray}
where $r=\sqrt{y^{2}_{1}+y^{2}_{2}+y^{2}_{3}}$  and
\begin{eqnarray}
&&v_{+}(y)=\left(\begin{array}{c}
            \cos\frac{\theta}{2}e^{-i\varphi}\\
            \sin\frac{\theta}{2}
            \end{array}\right), \nonumber\\ 
&&v_{-}(y)=\left(\begin{array}{c}
            \sin\frac{\theta}{2}e^{-i\varphi}\\
            -\cos\frac{\theta}{2}
            \end{array}\right)
\end{eqnarray}
by using the polar coordinates, 
$y_{1}=r\sin\theta\cos\varphi$, $y_{2}=r\sin\theta\sin\varphi$,
$y_{3}=r\cos\theta$.
If one defines
\begin{eqnarray} 
v^{\dagger}_{m}(y)i\frac{\partial}{\partial t}v_{n}(y)
=A_{mn}^{k}(y)\dot{y}_{k}
\end{eqnarray}
where $m$ and $n$ run over $\pm$,
we have
\begin{eqnarray}
A_{++}^{k}(y)\dot{y}_{k}
&=&\frac{(1+\cos\theta)}{2}\dot{\varphi}
\nonumber\\
A_{+-}^{k}(y)\dot{y}_{k}
&=&\frac{\sin\theta}{2}\dot{\varphi}+\frac{i}{2}\dot{\theta}
=(A_{-+}^{k}(y)\dot{y}_{k})^{\star}
,\nonumber\\
A_{--}^{k}(y)\dot{y}_{k}
&=&\frac{1-\cos\theta}{2}\dot{\varphi}.
\end{eqnarray}
The effective Hamiltonian corresponding to the Lagrangian (11)
is then given by 
\begin{eqnarray}
\hat{H}_{eff}(t)
&&=(E(0)+y_{0}(t) + g r(t))\hat{b}^{\dagger}_{+}
\hat{b}_{+}
+(E(0)+y_{0}(t) - g r(t))\hat{b}^{\dagger}_{-}\hat{b}_{-}
\nonumber\\
&& -\hbar \sum_{m,n}\hat{b}^{\dagger}_{m}A^{k}_{mn}(y)\dot{y}_{k}
\hat{b}_{n}
\end{eqnarray}
which is {\em exact} in the present two-level truncation.

In the conventional adiabatic approximation, one approximates
the effective Hamiltonian by
\begin{eqnarray}
\hat{H}_{eff}(t)&\simeq&
(E(0)+y_{0}(t) + g r(t))
\hat{b}^{\dagger}_{+}\hat{b}_{+}+(E(0)+y_{0}(t) - g r(t))
\hat{b}^{\dagger}_{-}\hat{b}_{-}
\nonumber\\
&&-\hbar [\hat{b}^{\dagger}_{+}A^{k}_{++}(y)\dot{y}_{k}
\hat{b}_{+}
+\hat{b}^{\dagger}_{-}A^{k}_{--}(y)\dot{y}_{k}\hat{b}_{-}]
\end{eqnarray}
which is valid for 
\begin{eqnarray}
Tg r(t)\gg \hbar\pi,
\end{eqnarray}
where $\hbar\pi$ stands for the magnitude of the geometric term 
times $T$.
The amplitude 
$\langle 0|\hat{\psi}(T)\hat{b}^{\dagger}_{-}(0)|0\rangle$, 
which corresponds to the probability amplitude in the first 
quantization, is given by
\begin{eqnarray}
\psi_{-}(T)
&\equiv&\langle 0|\hat{\psi}(T)\hat{b}^{\dagger}_{-}(0)|0\rangle\\
&=&\exp\{-\frac{i}{\hbar}\int_{0}^{T}dt[
E(0)+y_{0}(t) - g r(t) 
-\hbar A^{k}_{--}(y)\dot{y}_{k}] \} v_{-}(y(T))\nonumber
\end{eqnarray}
For a $2\pi$ rotation in $\varphi$ with fixed $\theta$, for 
example, the gauge invariant quantity gives rise to  
\begin{eqnarray}
\psi_{-}(0)^{\star}\psi_{-}(T)
&=&v_{-}(y(0))^{\star}v_{-}(y(T))
\nonumber\\
&&\times
\exp\{-\frac{i}{\hbar}\int_{0}^{T}dt[E(0)+y_{0}(t) - g r(t) 
-\hbar A^{k}_{--}(y)\dot{y}_{k}] \}\nonumber\\
&=&\exp\{i\pi(1-\cos\theta) \}\nonumber\\
&&\times \exp\{-\frac{i}{\hbar}\int_{C_{1}(0\rightarrow T)}dt
[E(0)+y_{0}(t) - g r(t)] \}
\end{eqnarray}
by using $v_{-}(y(T))=v_{-}(y(0))$ in the present 
choice of gauge, and the path $C_{1}(0\rightarrow T)$
specifies the integration along the above specific closed path.

The first phase factor $\exp\{i\pi(1-\cos\theta) \}$
stands for the familiar Berry's phase and the 
second phase factor stands for the conventional dynamical 
phase. The phase factor
is still invariant under a smaller set of gauge transformations
with
\begin{eqnarray}
\alpha_{-}(T)=\alpha_{-}(0)
\end{eqnarray}
and, in particular, for the gauge parameter of the form 
$\alpha_{-}(y(t))$.\\

\normalsize
\begin{figure}[!htb]
 \begin{center}
    \includegraphics[width=10.9cm]{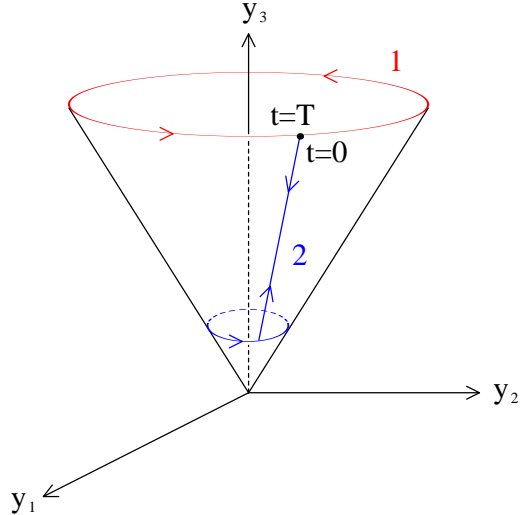} 
       \end{center}
\vspace{-9mm}      
 \caption{ %(Color online) 
 The path 1 gives the conventional 
geometric phase for a fixed finite $T$, 
whereas the path 2 gives a trivial geometric phase for a fixed finite $T$. Note that both of the paths cover the 
same solid angle $2\pi(1-\cos\theta)$.  }
\end{figure}

\vspace{1mm}

By deforming the path 1 to the path 2 in the parameter space in 
Fig. 1, it is shown that the amplitude (40) is replaced 
by~\cite{fujikawa, fujikawa2} 
\begin{eqnarray}
&&\psi_{-}(0)^{\star}\psi_{-}(T)=\exp\{-\frac{i}{\hbar}\int_{C_{2}(0\rightarrow T)}
dt[E(0)+y_{0}(t)-gr(t)] \}
\end{eqnarray}
 The path $C_{2}(0\rightarrow T)$ specifies the path 2 in 
Fig.1, and $v_{-}(y(T))=v_{-}(y(0))$ in the present choice of 
the gauge. Thus no geometric phase for the path $C_{2}$ for any 
fixed finite $T$. 

For $t=0$ or $t=T$, we start or end with the parameter 
region where the condition for the adiabatic 
approximation is satisfied. But approaching
the infinitesimal neighborhood of the origin where the level 
crossing takes place, the condition (38) is no more satisfied and
instead one has $Tgr\ll \hbar$.
In this region of the parameter space, $\hat{H}_{eff}$ 
is replaced by 
\begin{eqnarray}
\hat{H}_{eff}(t)&\simeq& (E(0)+y_{0}(t))
\hat{c}^{\dagger}_{+}\hat{c}_{+}
+(E(0)+y_{0}(t))\hat{c}^{\dagger}_{-}\hat{c}_{-}
-\hbar\dot{\varphi} \hat{c}^{\dagger}_{+}\hat{c}_{+}
\end{eqnarray}
by performing a unitary transformation to a regular basis
\begin{eqnarray}
\hat{b}_{m}=\sum_{n}U(\theta(t))_{mn}\hat{c}_{n}
\end{eqnarray}
with
\begin{eqnarray}
U(\theta(t))=\left(\begin{array}{cc}
            \cos\frac{\theta}{2}&-\sin\frac{\theta}{2}\\
            \sin\frac{\theta}{2}&\cos\frac{\theta}{2}
            \end{array}\right)
\end{eqnarray}

We thus conclude that the topological 
interpretation of the Berry's phase fails in the practical 
Born-Oppenheimer approximation where $T$ is identified with the 
period of the slower dynamical system.

\section{Gauge symmetry in non-adiabatic phase}

We briefly comment on the gauge symmetry invloved in the 
non-adiabatic phase proposed by Aharonov and 
Anandan~\cite{aharonov}.
The analysis starts with the wave function satisfying 
\begin{eqnarray}
&&\int d^{3}x \psi(t,\vec{x})^{\star}\psi(t,\vec{x})=1,
\end{eqnarray}
and  a cyclic condition
\begin{eqnarray}
&&\psi(T,\vec{x})=e^{i\phi}\psi(0,\vec{x})
\end{eqnarray}
with a real constant $\phi$. 
These properties then imply the existence of a hermitian 
Hamiltonian
\begin{eqnarray}
i\hbar\frac{\partial}{\partial t}\psi(t,\vec{x})=
\hat{H}(t,\frac{\hbar}{i}\frac{\partial}{\partial\vec{x}},
\vec{x})\psi(t,\vec{x})
\end{eqnarray}
but the motion of $X(t)$ is arbitrary.

The analysis of non-adiabatic phases is based on the equivalence class which identifies
 all the vectors of the form 
\begin{eqnarray}
\{e^{i\alpha(t)}\psi(t,\vec{x})\}.
\end{eqnarray}
The conventional Schr\"{o}dinger equation is not invariant
under this equivalence class, we may thus define an equivalence 
class of Hamiltonians
\begin{eqnarray}
\{ \hat{H} -\hbar\frac{\partial}{\partial t}\alpha(t)\}.
\end{eqnarray}
We next define an object~\cite{aitchison, mukunda} 
\begin{eqnarray}
\Psi(t,\vec{x})\equiv\exp[i\int_{0}^{t}dt \int d^{3}x
\psi(t,\vec{x})^{\star}
i\frac{\partial}{\partial t}\psi(t,\vec{x}) \psi(t,\vec{x})
\end{eqnarray}
which satisfies
\begin{eqnarray}
&&\Psi(0,\vec{x})=\psi(0,\vec{x}),\nonumber\\
&&\int d^{3}x \Psi(t,\vec{x})^{\star}
i\frac{\partial}{\partial t}\Psi(t,\vec{x})=0.
\end{eqnarray}
Under the equivalence class transformation (or gauge 
transfromation)
\begin{eqnarray}
\psi(t,\vec{x}) \rightarrow e^{i\alpha(t)}\psi(t,\vec{x}),
\end{eqnarray}
$\Psi(t,\vec{x})$ transforms as 
\begin{eqnarray}
\Psi(t,\vec{x}) \rightarrow e^{\alpha(0)}
\Psi(t,\vec{x}).
\end{eqnarray}

The gauge invariant quantity is then defined by
\begin{eqnarray}
\Psi(0,\vec{x})^{\star}\Psi(T,\vec{x})=\psi(0,\vec{x})^{\star}\exp[i\int_{0}^{T}dt \int d^{3}x
\psi(t,\vec{x})^{\star}
i\frac{\partial}{\partial t}\psi(t,\vec{x}) ]
\psi(T,\vec{x})
\end{eqnarray}
By a suitable gauge transformation 
\begin{eqnarray}
\psi(t,\vec{x})\rightarrow \tilde{\psi}(t,\vec{x})=
e^{-i\alpha(t)}\psi(t,\vec{x})
\end{eqnarray}
with
\begin{eqnarray}
\alpha(T)-\alpha(0)=\phi
\end{eqnarray}
we can make the prefactor 
\begin{eqnarray}
\psi(0,\vec{x})^{\star}\psi(T,\vec{x})\rightarrow
\tilde{\psi}(0,\vec{x})^{\star}\tilde{\psi}(T,\vec{x})&=&e^{i\alpha(0)}\psi(0,\vec{x})^{\star}e^{-i\alpha(T)}
\psi(T,\vec{x})\nonumber\\
&=&|\psi(0,\vec{x})|^{2}
\end{eqnarray} 
real and positive for a cyclic evolution.
The above gauge invariant quantity is then given by 
\begin{eqnarray}
\Psi(0,\vec{x})^{\star}\Psi(T,\vec{x})
=|\psi(0,\vec{x})|^{2}\exp[i\int_{0}^{T}dt \int d^{3}x
\tilde{\psi}(t,\vec{x})^{\star}
i\frac{\partial}{\partial t}\tilde{\psi}(t,\vec{x}) ]
\end{eqnarray}
and the factor on the exponential extracts all the information
about the phase from the gauge invariant quantity.

The "non-adiabtic" phase is then defined by 
\begin{eqnarray}
\beta=\oint dt \int d^{3}x
\tilde{\psi}(t,\vec{x})^{\star}
i\frac{\partial}{\partial t}\tilde{\psi}(t,\vec{x})
\end{eqnarray}
with
\begin{eqnarray}
\tilde{\psi}(0,\vec{x})=\tilde{\psi}(T,\vec{x})
\end{eqnarray}
This phase describes certain intrinsic properties of the 
Schr\"{o}dinger equation, and it is invariant under a residual 
gauge symmetry with 
\begin{eqnarray}
\alpha(T)=\alpha(0).
\end{eqnarray}
We emphasize that 
the gauge symmetry here is quite different from our hidden local
symmetry which is related to an arbitrariness of the choice of 
coordinates in the functional space.
\\

The geometric phases in classical mechanics have also been 
analyzed in the literature~\cite{review}.

\end{document}